\documentclass[12pt,preprint]{aastex}
\usepackage{amsmath}

\begin{document}
\title{Fermi Constrains Dark Matter Origin of High Energy
Positron Anomaly}
\author{Martin Pohl}
\affil{Institut f\"ur Physik und Astronomie, Universit\"at
Potsdam, 14476 Potsdam-Golm, Germany}
\affil{DESY, 15738 Zeuthen, Germany}
\email{pohlmadq@gmail.com}
\author{David Eichler}
\affil{Physics Department, Ben-Gurion University,
Beer-Sheva 84105, Israel}
\email{eichler@bgumail.bgu.ac.il}

\begin{abstract}
Fermi measurements of the high-latitude $\gamma$-ray background strongly constrain a decaying-dark-matter origin for the 1--100 GeV
Galactic positron anomaly measured with PAMELA. Inverse-Compton scattering of the microwave background
%in the outer Milky-Way halo and  in intergalactic space
by the emergent positrons produces a bump in the diffuse 100-200 MeV
$\gamma$-ray background that would protrude from the observed background at these
energies. The positrons are thus constrained to emerge from
the decay process at a typical energy between $\sim$100 GeV and $\sim$250 GeV. By considering
only $\gamma$-ray emission of the excess positrons and electrons, we derive a minimum diffuse $\gamma$-ray flux that, apart from the positron spectrum assumed, is independent of the actual decay modes.
Any $\gamma$-rays produced directly by  the dark-matter decay leads to
an additional signal that make the observational limits more severe. A similar
constraint on the energy of emergent positrons from annihilation in dark-matter substructures is argued to exist, according to recent estimates of enhancement in low-mass
dark-matter substructures, and improved simulations of such substructure will further sharpen this constraint.
\end{abstract}

\section{Introduction}

The positron excess in Galactic cosmic ray positrons between 1 and
$100$ GeV, despite a history of conflicting results, appears to be
confirmed by PAMELA \citep{pam}. There are, of course, several possible
astrophysical explanations.  Nearby sources (e.g. supernova or pulsars) of positrons, for example,  which suffer fewer losses than typical Galactic  positrons because they are younger \citep[e.g.][]{ME05, profumo08}, contribute harder spectra. Nevertheless,   dark matter
annihilation \citep[e.g.][]{TE87,Tylka1989,hooper09} or decay
\citep{Eichler1989,Arvanitaki} have long been suggested as a possible source.

Annihilation, however, encounters a number of difficulties, as it
requires a substantial boosting either from clumping
\citep[e.g.][]{km09} or from a Sommerfeld enhancement
\citep{Arkani-Hamed}, and would lead to intense photon emission from
the Galactic-Center region \citep[e.g.][]{ME05,zhang} or
extragalactic background radiation from the superposition of haloes
\citep{profumo09}, that appears to exceed current observational
limits.

Here we investigate dark-matter decay as the source of the excess
positrons observed with PAMELA. Our purpose is not a comprehensive theory of the
dark-matter decay, in particular not the viability of leptophilic decay, which is one
of the challenges \citep{Arkani-Hamed},{ given} that standard
hadronization scenarios are already excluded by the lack of an excess in the
cosmic-ray antiproton data \citep{adriani1}.
The decay source function scales with the
dark-matter density, not the density squared as does annihilation, and therefore
the limits on {decaying} dark matter from observations of high-energy $\gamma$ rays from,
e.g., dwarf galaxies are weak \citep{essig}, but diffuse emission may provide
more stringent constraints \citep{ishiwata,chen}. We calculate the diffuse galactic
$\gamma$-ray emission of the excess leptons, presuming they arise from dark-matter decay. The
$\gamma$-ray intensity thus derived is more model-independent and also a lower limit to
the true emission level, because we
do not count $\gamma$ rays that are directly produced in the dark-matter decay,
possibly via other unstable particles.

\section{Positron production by dark-matter decay}
\subsection{The differential number density of positrons}
The propagation length of positrons and electrons at energies $\gtrsim 50$~GeV is
$L_p\simeq 2\sqrt{\kappa\,\tau_{\rm loss}}\lesssim 600$~pc \citep{p03,lat-int}, considerably
less than the scalelength of the dark-matter halo in the Galaxy, i.e. the source
distribution of positrons and electrons from dark-matter decay. Here, $\kappa$ is the energy-dependent
diffusion coefficient and $\tau_{\rm loss}$ is the electron/positron energy-loss timescale.
To first order, the spatial
redistribution of those electrons by diffusive and convective transport can therefore be neglected.
The differential number density of positrons has, in good approximation,  the same spatial profile as the
dark-matter density, $\rho_{\rm DM}$, because in decay processes the positron source function obeys
$Q_{e^+}\propto \rho_{\rm DM}$. The differential number density of electrons/positrons in
the steady state is then given by
\begin{equation}
N(E, {\bf x})= \frac{1}{\vert {\dot E}(E)\vert} \int_E du\ Q_{e^+/e^-} (u)
\label{eq1}
\end{equation}
Above 10 GeV electron energy, the electron energy-loss rate, ${\dot E}(E)$, is
dominated by synchrotron and inverse Compton losses. The soft-photon
density varies somewhat between the Galactic plane and the halo, reaching a
maximum of 1.5~eV/cm$^3$ about 300 pc above the midplane and falling to
0.8~eV/cm$^3$ at 5 kpc above the midplane \citep{porter}. Here we use an average value
$U_{\rm ph} =1.1\ {\rm eV/cm^3}$ to calculate the energy losses. The
magnetic-field strength is not well known, and, using
a canonical constant value for it, $8\ {\rm \mu
G}$, we find for the energy-loss rate
\begin{equation}
{\dot E}(E)=-\frac{E}{\tau_{\rm loss}}\simeq - (2.5\cdot 10^{-14}\ {\rm GeV\,s^{-1}})\,\zeta\,
\left(\frac{E}{10\ \rm GeV}\right)^2\, ,
\label{eq2}
\end{equation}
where we ignore the weak modifications arising from the Klein-Nishina corrections
to the inverse-Compton
scattering rate. The parameter $\zeta$ permits a scaling of the energy-loss
rate relative to that given by the model for the galactic radiation field and a
magnetic-field strength $B=8\ {\rm \mu G}$.
\begin{equation}
\zeta=\frac{U_{\rm ph}+ \frac{B^2}{8\pi}}{2.7\ {\rm eV/cm^3}}
\label{eq2a}
\end{equation}
The microwave background implies
a lower limit $\zeta \ge 0.1$, and a magnetic-field strength $B=5\ {\rm \mu G}$
would be described by $\zeta=0.64$. Estimates of the Galactic magnetic field
based on modeling the radio synchrotron emission give
amplitudes from $B=6.1\ {\rm \mu G}$ \citep{strong00} to
$B=10\ {\rm \mu G}$ \citep{pe98}, and the most likely value for $\zeta$ is
therefore in the range $0.75\le\zeta\le 1.3$.
The differential number
density of electrons/positrons cannot have a spectrum
harder than $E^{-2}$, which is the spectrum that comes from the hardest possible injection spectrum.

The PAMELA collaboration has measured a positron fraction that is rapidly increasing above 10~GeV. The
highest-energy data point is
\begin{equation}
\frac{N_{e^+}}{N_{e^+}+N_{e^-}}\,(82\ {\rm GeV})=0.137 \pm 0.045
\label{eq3}
\end{equation}
about 90\% of which is in excess to the positrons expected from secondary
production in cosmic-ray interactions. The total electron spectrum has not been
determined with PAMELA yet, but we may use the excellent Fermi-LAT
data that between 20~GeV and 1~TeV are well represented by a single power law \citep{lat-dat}
\begin{equation}
N_{e^+}+N_{e^-}=(73.5 \pm 2.6)\cdot 10^{-7}\,\left(\frac{E}{\rm GeV}\right)^{-(3.045 \pm 0.008)}\
{\rm GeV^{-1}\,m^{-3}}
\label{eq4}
\end{equation}
Combining Eqs.~\ref{eq3} and \ref{eq4} we find the positron density at 82 GeV as
\begin{equation}
N_{e^+} (82\ {\rm GeV})\simeq 1.5\cdot 10^{-12} \ {\rm GeV^{-1}\,m^{-3}}
\label{eq5}
\end{equation}
The positron excess over ordinary secondary
production is measured to be only about 0.03 at 17~GeV, thus indicating that
the excess component has a spectrum not significantly softer than $E^{-2}$.
Equation~\ref{eq1} indicates that this requires the injection of electron and positrons
with a typical energy above 80 GeV. For simplicity we assume the injection of
monoenergetic electrons,
\begin{equation}
Q_{e^+/e^-}=Q_0\, \delta\left(E-E_{\rm max}\right)
\label{eq-inj}
\end{equation}
where $E_{\rm max}\lesssim 500$~GeV to avoid a spectral feature
in the total electron spectrum, which would be in conflict with the
power-law fit to the Fermi data. We shall see that the $\gamma$-ray limits 
calculated below provide much tighter constraints on $E_{\rm max}$.
Assuming an equal number of excess electrons
and positrons, we therefore estimate the total differential
density of electrons/positrons that may come from dark-matter decay as
\begin{equation}
N(E, {\bf x})= (2.2\cdot 10^{-8} \ {\rm GeV^{-1}\,m^{-3}})\, R({\bf x})\,
\left(\frac{E}{\rm GeV}\right)^{-2} \,\Theta\left(E_{\rm max} -E\right)
\label{eq6}
\end{equation}
\begin{figure}
  \centering
  \includegraphics[width=3in]{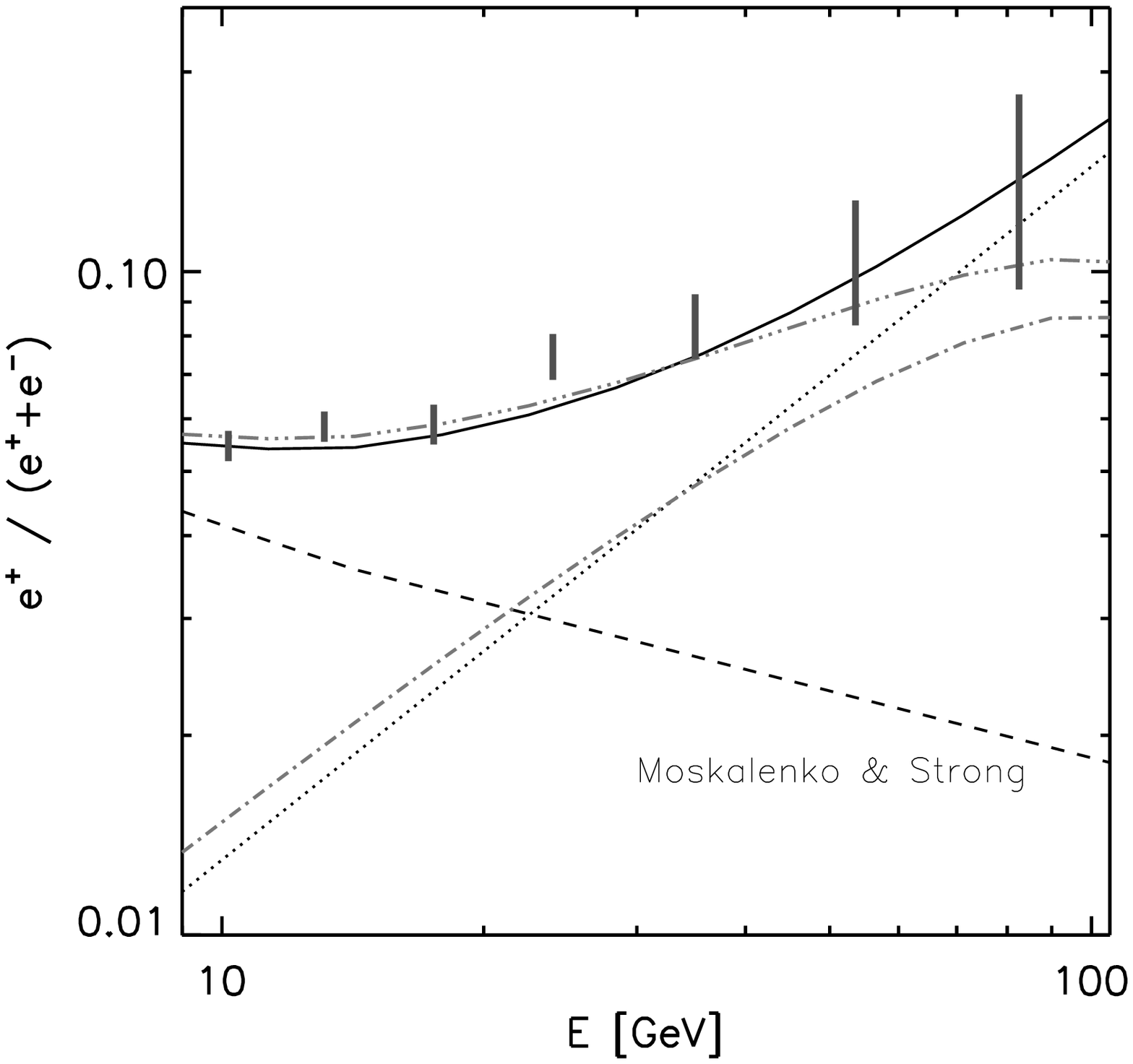}
  \caption{The measured positron fraction between 10 GeV and 82 GeV shown in comparison
with a calculation for pure secondary production of positrons \citep{ms98}, the
dark-matter contribution according to Eq.~\ref{eq6} (dotted line), and the total
calculated positron fraction (solid line). For comparison, we also show the positron 
fraction for a flat injection spectrum up to 200~GeV (dot-dashed line) and the total positron fraction for that case (triple-dot-dashed line).}
  \label{fig2}
\end{figure}
In Figure~\ref{fig2} we demonstrate that the sum of this modeled dark-matter
decay component and the expected contribution of secondary positron production
provides a good fit to the PAMELA data. For comparison, the figure also displays
the positron fraction for the case of a flat injection spectrum, which may result
from the decay of an intermediate particle with high kinetic energy.
 \begin{equation}
Q_{\rm alt.}=\frac{Q_0}{E_0}\, \Theta\left(E_{\rm max}-E\right)
\label{eq-inj-alt}
\end{equation}
\begin{equation}
\Rightarrow \ N_{\rm alt.} (E, {\bf x})\propto 
R({\bf x})\,\left(1-\frac{E}{E_{\rm max}}\right)\,
\left(\frac{E}{\rm GeV}\right)^{-2} \,\Theta\left(E_{\rm max} -E\right)
\label{eq6alt}
\end{equation}
Here $Q_0$ is chosen 20\% larger than in the case of monoenergetic injection to 
improve the fit of the positron fraction. To be noted from the figure is that
$E_{\rm max}\gtrsim 200\ {\rm GeV}$ is needed for flat injection to well 
reproduce the positron fraction measured with PAMELA.

$R({\bf x})$ denotes the spatial
distribution normalized to the local density, which should reflect
that of the dark-matter particles. Simulations suggest that the
dark-matter halo in the Galaxy is not spherical, and the vertical
scale height may be only 40\% of that in the Galactic plane
\citep{kuhlen07}. The tidal streams produced by the Sagittarius
dwarf speroidal may even indicate a triaxial density distribution
\citep{law09}. Here we conservatively assume a dark-matter halo with vertical
flattening and for simplicity a global $r^{-2}$ density
scaling instead of, e.g., a NFW halo \citep{navarro}. In that
approximation, the normalized density profile at the solar circle
($r=R_0=8$~kpc) is
\begin{equation}
R({\bf x})=\frac{\rho(R_0,z)}{\rho(R_0,z=0)}\simeq \left(1+2.5\,\frac{z^2}{R_0^2}\right)^{-1}
\label{eq7}
\end{equation}

\subsection{The dark-matter lifetime}
Using Eqs.~\ref{eq1}, \ref{eq2}, and \ref{eq6} we can estimate the source
function of pairs from dark matter,
and hence the dark-matter decay timescale.
Assuming for simplicity decay into monoenergetic pairs, we find from Equations
\ref{eq1} and \ref{eq-inj}
\begin{equation}
Q_0=\zeta\ (5.5\cdot 10^{-30}\ {\rm cm^{-3}\,s^{-1}})
\label{eq-source}
\end{equation}
The source power is
\begin{equation}
P=\int dE\ E\,Q_{e^+/e^-}  =Q_0\,E_{\rm max}=\zeta\
(1.1\cdot 10^{-27}\ {\rm GeV\,cm^{-3}\,s^{-1}})\,
\left(\frac{E_{\rm max}}{200\ \rm GeV}\right)
\label{eq-power}
\end{equation}
Comparing with the local dark-matter mass density, $\rho_0\simeq 0.3\ {\rm GeV\,cm^{-3}}$, and
accounting for an arbitrary branching ratio
of conversion into pairs, $\eta$, we find for the dark-matter decay time
\begin{equation}
\tau_{\rm decay}=\frac{\rho_0\,\eta}{P}\simeq
\frac{\eta}{\zeta}\,(2.7\cdot 10^{26}\ {\rm s})\,
\left(\frac{E_{\rm max}}{200\ \rm GeV}\right)^{-1}
\label{eq-time}
\end{equation}
Care must be exercised in comparing astrophysical limits on the decay time on
account of its dependence on the energy of the injected pairs and on the assumed
magnetic-field strength. For example, \citet{chen} use $B=3\,{\rm \mu G}$, implying
$\zeta=0.5$, and consequently their limits on $\tau_{\rm decay}$ are relatively high.

\section{Inverse Compton scattering}
\subsection{Galactic emission}
The electrons and positrons in the halo will inverse-Compton scatter soft photons
into the $\gamma$-ray
band where they can be observed with, e.g., the Fermi-LAT detector. The target
photon field includes
the microwave background, galactic infrared emission, and galactic optical emission,
only the first of which
is isotropic. The latter two will be backscattered toward the Galaxy, thus somewhat
increasing the scattering
rate compared with the isotropic case. The radiation field has been recently modeled
out to $z=5$~kpc \citep{porter}. Surprisingly, the energy density of optical light is higher
in the halo than in the midplane on account of the thin-disk distribution of absorbers.

The differential cross section for the up-scattering of isotropic photons of energy
$\epsilon$ to energy $E_\gamma$ is given by \citep{bg70}
\begin{equation}
\sigma(E_\gamma,\epsilon,E)= \frac{3\,\sigma_T\,(m_e\,c^2)^2}{4\,\epsilon\,E^2}
\,\left[2\,q\,\ln q+1+q-2\,q^2+\frac{(1-q)\,q^2\,\Gamma_e^2}{2\,(1+q\,\Gamma_e)}\right]
\label{eq8}
\end{equation}
where
\begin{equation}
q=\frac{E_\gamma}{\Gamma_e\,(E-E_\gamma)}\quad {\rm and}\ \Gamma_e=\frac{4\,\epsilon
\,E}{m_e^2\,c^4}
\label{eq9}
\end{equation}
The $\gamma$-ray intensity observed from the Galactic pole is then computed using
the differential density of electrons/positrons (Eq.~\ref{eq6}) and the
differential density of soft photons, $n(\epsilon)$ \citep{porter}.
\begin{equation}
I_\gamma=\frac{c}{4\pi}\,\int_0 dz\ \int dE\ N(E, {\bf x})\,\int d\epsilon\
n(\epsilon)\,\sigma(E_\gamma,\epsilon,E)
\label{eq10}
\end{equation}
The resulting intensity is shown in Figure~\ref{fig1}. To be noted from the figure is that even at
30 GeV, where the emission arises from upscattering optical photons, the expected intensity is
below 20\% of the preliminary estimate of the extragalactic isotropic background observed with
Fermi \citep{ackermann}, in line with the findings of \citet{ishiwata}.
Diffuse galactic emission from cosmic-ray interaction has an intensity
similar to that of the extragalactic isotropic background, and hence dark-matter decay is poorly
constrained by $\gamma$-ray emission in the Galaxy.

Whereas no significant intensity above 1 GeV is expected to come from beyond a distance of
5 kpc, the limit to which we have integrated the galactic emission, this is not true for
the upscattering of microwave-background photons into the 100-MeV band. In fact, in the outer
halo inverse-Compton scattering of the microwave background accounts for a larger share of
the electron energy losses than near the Galactic plane, because the infrared and optical
photon fields quickly lose intensity beyond 5 kpc above the plane of the Galaxy. Also,
the magnetic-field strength is expected to fall off, although we do not know at what
point it drops to 3$\ {\rm \mu G}$, below which the synchrotron energy losses are
subdominant. We will estimate the intensity in the 100-MeV band from galactic
dark-matter decay in the outer halo in the next section, together with
the extragalactic component.

\subsection{Extragalactic background from intergalactic dark matter}
\begin{figure}
  \centering
  \includegraphics[width=3in]{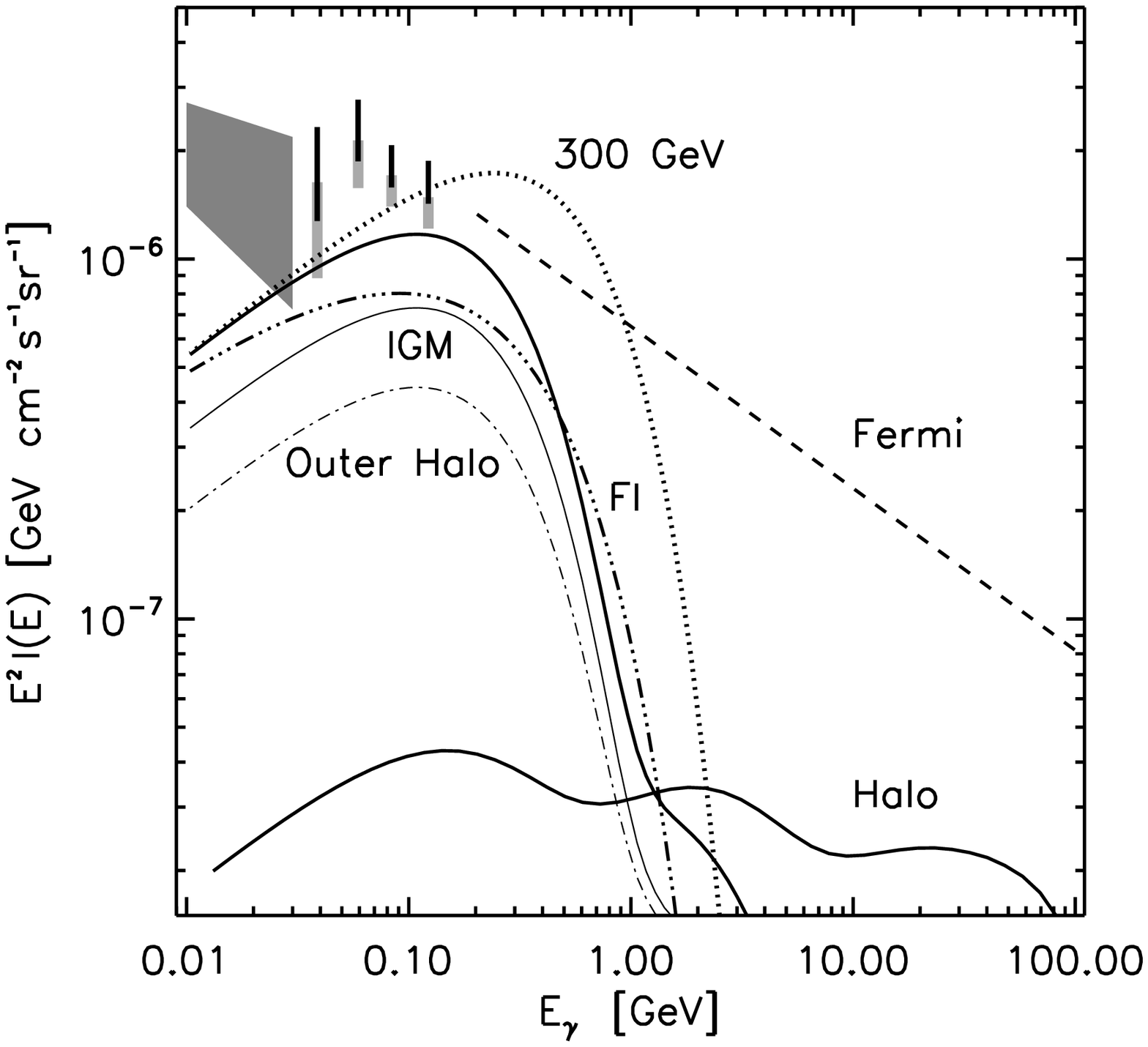}
  \caption{Comparison of the predicted $\gamma$-ray intensity observed from the galactic poles
with the preliminary estimate of the extragalactic isotropic background observed with Fermi,
indicated by the dashed line \citep{ackermann}, and older data from COMPTEL
\citep{kapp} and EGRET \citep{sreekumar,smr05}. The thick line labeled Halo shows galactic emission
according to equation~\ref{eq10}, and the other lines are for $\gamma$-ray
emission from intergalactic dark matter, all for $\zeta=1$ and except where noted for
$E_{\rm max}=200\ {\rm GeV}$. In detail, the thin solid line is
extragalactic emission as given in equation~\ref{eq14}, the dot-dashed line is
emission from the outer Milky-Way halo beyond $z=10\ {\rm kpc}$, the thick solid
line the sum of the two, and the dotted line is the same for $E_{\rm max}=
300\ {\rm GeV}$. The triple-dot-dashed line labeled FI is for flat injection 
up to 300~GeV (cf. Eqs. \ref{eq-inj-alt} and \ref{eq6alt}).}
  \label{fig1}
\end{figure}

The majority of dark matter is located sufficiently far away from galaxies ($\gtrsim 20$~kpc)
that positrons and electrons from its decay would primarily interact with the microwave background.
Since all electrons suffer the same fate and the electron source rate scales linearly with the
dark-matter density, we can ignore any density structure and use spatially averaged quantities,
i.e. a conservative fraction 90\% of $\Omega_m=0.239$ times the critical density,
$9.9\cdot 10^{-30}\ {\rm g\,cm^{-3}}$.
For the source function and the energy-loss rate we find
\begin{equation}
Q_{e^+/e^-}\simeq (1+z)^3\,\zeta\,(2.2\cdot 10^{-35}
\ {\rm cm^{-3}\,s^{-1}})\,
\delta\left(E-E_{\rm max}\right)
\label{eq11}
\end{equation}
and
\begin{equation}
{\dot E}= (1+z)^4 (-2.5\cdot 10^{-17}\ {\rm GeV\,s^{-1}})\,
\left(\frac{E}{\rm GeV}\right)^2
\label{eq12}
\end{equation}
where $z$ denotes the redshift.
The average differential number density is therefore
\begin{equation}
N(E,z)= \frac{\zeta}{1+z}\,(8.8\cdot 10^{-19} \ {\rm GeV^{-1}\,cm^{-3}})\,
\left(\frac{E}{\rm GeV}\right)^{-2} \,\Theta\left(E_{\rm max} -E\right)
\label{eq13}
\end{equation}
The intensity of inverse-Compton emission is calculated as
\begin{equation}
I_\gamma=\frac{c}{4\pi\,(1+z)^2}\,\int_0 dz\ f(z)\,\int dE\ N(E,z)\,\int d\epsilon\
n(\epsilon)\,\sigma\left[E_\gamma\,(1+z),\epsilon,E\right]
\label{eq14}
\end{equation}
where
\begin{equation}
f(z)=\frac{c}{H_0\,(1+z)}\,\left[(1+z)^2\,(1+\Omega_m\,z)-z\,(2+z)\,
\Omega_\Lambda\right]^{-\frac{1}{2}}
\label{eq15}
\end{equation}
is the Jacobian used to turn the line-of-sight integral into a redshift integral. We integrate
to redshift $z=5$, beyond which the emissivity is negligibly small.
The soft-photon distribution is the cosmological microwave background
\begin{equation}
n(\epsilon)=\frac{8\pi\,\epsilon^2}{h^3\,c^3}\,
\frac{1}{\exp\left(\frac{\epsilon}{\epsilon_0\,(1+z)}\right)-1}
\label{eq16}
\end{equation}
with $\epsilon_0=2.4\cdot 10^{-4}$~eV, added to which is a small contribution in the
infrared band \citep{franceschini}.
The resulting expected $\gamma$-ray intensity
is plotted in Figure~\ref{fig1} for $\zeta=1$, together with that produced in the outer parts
of the Milky-Way halo at height $z\ge 10$~kpc where IC scattering off the CMB is also dominant and the 
expected differential density of electrons and positrons from dark-matter decay is
\begin{equation}
N(E, {\bf x})= (2.4\cdot 10^{-7} \ {\rm GeV^{-1}\,m^{-3}})\, R({\bf x})\,
\left(\frac{E}{\rm GeV}\right)^{-2} \,\Theta\left(E_{\rm max} -E\right)\ ,
\label{eq6neu}
\end{equation}
larger than the galactic-plane solution (Eq.~\ref{eq6}) by the ratio of the energy-loss lifetime in
the outer halo to that in the galactic plane. 
If the excess electrons and positrons could freely stream in the outer halo, we would be
overestimating their $\gamma$-ray emission which contributes at most 1/3 of the total
dark-matter signal. However, a number of well-known plasma instabilities would sharply impede such
free streaming. As a
100-GeV electron has a propagation length of only $\sim$15~kpc even if the diffusion coefficient is a
100 times that in the galactic plane, the effect of particle propagation is neglected
in deriving Eq.~\ref{eq6neu}.

At 200~MeV $\gamma$-ray energy, the predicted intensity is close to that observed
with Fermi \citep{ackermann}. The uncertainty in the measured intensity is
typically 15\% and predominantly systematic in origin.
Because the electron/positron density is linear in $\zeta$, the predicted
$\gamma$-ray intensity would exceed the observational limits if the total
magnetic-field in the solar vicinity were stronger than $10\ {\rm \mu G}$. It would also exceed
the observational limits if the characteristic energy of the injected pairs were higher
than $\sim$250~GeV. For comparison, Figure~\ref{fig1} also shows the expected intensity for
$E_{\rm max}=300\ {\rm GeV}$ to be twice that observed at 200~MeV $\gamma$-ray energy.

The same scaling with energy $E_{\rm max}$ results for flat injection
(cf. Eqs. \ref{eq-inj-alt} and \ref{eq6alt}). The $\gamma$-ray peak will appear at slightly lower 
energy and with somewhat reduced flux compared with monoenergetic injection at   
the same $E_{\rm max}$, because the mean particle energy is $E_{\rm max}/2$.
This is demonstrated as well in Figure~\ref{fig1} where we show the expected 
$\gamma$-ray bump resulting from flat injection up to 300~GeV, i.e. with
characteristic energy $\sim$~150~GeV.

\section{Conclusions}
We have investigated $\gamma$-ray constraints on the notion that dark-matter decay is responsible
for the recently measured excess of cosmic-ray positrons in the 1--100 GeV band.
%The $\gamma$-ray emission of the excess positrons and %electrons constrains their injection energy to be $\lesssim %200 $GeV, while the hard positron spectrum, consistent with %$E^{-2}$, constrains the injection spectrum to be above 80 %GeV. This is a very tight constraint that to our knowledge %rules out most if not all specific decay scenarios that %have been proposed thus far.
While the predicted GeV-band intensity from the inverse-Compton
scattering of infrared and optical photons is below a preliminary
estimate of the extragalactic $\gamma$-ray background based on Fermi
data \citep{ackermann}, the {\it extragalactic} background from the decay
of intergalactic dark matter would produce a bump at 100--300 MeV
that is close to the observed extragalactic background at these
energies, even  one estimated from the older EGRET data
\citep{sreekumar}. Dark-matter decay therefore does not seem to
be a viable explanation of the positron excess, unless the characteristic energy
of the pairs produced in dark-matter decay is only in a narrow window  between 
the lower limit $\sim 100\ {\rm GeV}$ imposed by high-energy limit of the Pamela 
measurement and the high energy limit of $\sim 250$ GeV imposed by the extragalactic 
100 to 200 MeV $\gamma$-ray background. 

Our stated limits are conservative in several ways: a) Any $\gamma$ rays directly produced
in the decay of the dark-matter particles will lead to
an additional signal that will make the observational limits more severe.
b) We have neglected the contribution of astrophysical sources that are needed to explain
the {\it shape} of  the observed $\gamma$ ray spectrum, which is far broader than the dark-matter
contribution.  Any dark matter contribution would probably be noticeable at the 30 percent level.
c) Any real decay process may involve multi-generational pairs, such as decay into electron
positron pairs via taus and muons, even if quarks are avoided by some leptophilic process.
This makes it harder to fit the pairs into the allowed energy window.

We stress that the predicted 100-200 MeV $\gamma$-ray intensity from
dark-matter decay is nearly model-independent, because it depends
only on the total dark-matter density in regions outside those of
strong galactic magnetic fields and starlight, i.e. in regions where
inverse-Compton scattering of microwave background photons is a
calorimeter for intergalactic electrons and positrons on account of
its dominance among the energy-loss processes. The only assumptions
we have made are time-independence of the dark-matter lifetime,
which follows from conventional elementary particle physics, and the
best estimates for the average magnetic field of the Galaxy in the
solar neighborhood.

Moreover, because the anomalous positrons reported by PAMELA are too
energetic to retain their energy for a Hubble time, the  only way to
prevent the energy from going into $\gamma$ rays is to preempt the
inverse Compton losses with even faster synchrotron losses. It would seem then that
magnetic fields of strength $\gtrsim 3\ \mu{\rm G}$ are the only way to 
preempt IC losses.

The argument would then apply to positrons from
clump-enhanced dark matter annihilation as well as decay, if such
enhancement occurred a) mostly in the outer part of galaxies where the
magnetic field and starlight are weak, b) in extragalactic dark matter clumps or dwarf galaxies,
where the magnetic field is weaker that in our own galaxy or c) elliptical galaxies, where the
magnetic field strengths are likely to be less than in our own Galaxy.
Quantifying  this  constraint for annihilating dark matter requires a
reliable computation of the distribution of annihilation, which is limited by numerical resolution
at small mass scales. \citet{diemand} argue that the positron emissivity per unit mass
(q) in the outer halo, $q_{oh}$,  exceeds that in the local galactic neighborhood, $q_{ln}$,
because substructure survives tidal destruction far better there. This would raise the
$\gamma$ ray contribution in the outer halo by a factor of $q_{oh}/q_{ln}$ relative to
the limits we derive for decaying dark matter, where $q_{oh}/q_{ln} =1$ everywhere.
Extragalactic substructure may give an even higher $q$, because there is less tidal
destruction, though there is the competing effect that in regions of lower density
fluctuation, substructure is less likely to form at high redshift.
Sommerfeld enhancement of the annihilation cross section
at low center-of-mass energy would further raise the contribution of low mass substructures, in
which the center-of-mass kinetic energy of the annihilating pairs would be lower than
in the diffuse
dark matter in the solar neighborhood \citep{KMS,KS}. Thus, if we could make 
reliable simulations of dark-matter
substructure down to the smallest mass scales for all cosmic locations, the results would probably
already rule out explaining the PAMELA positron excess by annihilating dark matter.

The arguments here do not, of course, immediately  rule out detection of decay or annihilating 
dark matter by other more sensitive means, e.g. $\gamma$ rays from localized DM concentrations. 
They merely argue against most scenarios for explaining the PAMELA positron anomaly in this 
manner. However, they call attention to the fact that any scenario for making $\gamma$ rays 
that includes accompanying e+ e- pairs, e.g.  $\gamma$ rays by pion decay or final state 
radiation, may be constrained by the inverse Compton $\gamma$ rays that would accompany 
the direct $\gamma$ rays emitted in regions of weak magnetic field. Conversely,  annihilation 
or decay that produces more  pairs than  direct $\gamma$ rays may be more readily detected 
via inverse Compton radiation of the pairs.

\acknowledgements We thank  M. Kuhlen and P. Madau for helpful discussions. We acknowledge 
support from the U.S.-Israel
Binational Science Foundation, the Israel Academy of Science, and
the Robert and Joan Arnow Chair of Theoretical Astrophysics. This
research was supported in part by the National Science Foundation
under Grant No. PHY05-51164.


\begin{thebibliography}{99}
\bibitem[Abdo et al.(2009)]{lat-dat} Abdo, A.A. et al. (The Fermi collaboration) 2009, 
PRL 102, 181101
\bibitem[Ackermann(2009)]{ackermann} Ackermann, M., for the Fermi collaboration, 2009, Talk at
International Cosmic Ray Conference, Lodz, Poland
\bibitem[Adriani et al.(2009a)]{pam} Adriani, O., Barbarino, G.C.,
Bazilevskaya, G.A., et al. 2009a, Nature 458, 607
\bibitem[Adriani et al.(2009b)]{adriani1} Adriani, O., Barbarino, G.C.,
Bazilevskaya, G.A., et al. 2009b, PRL 102, 051101
\bibitem[Arkani-Hamed et al.(2009)]{Arkani-Hamed} Arkani-Hamed, N., Finkbeiner,
D.P., Slatyer, T.R., Weiner, N. 2009, PRD 79, 015014
\bibitem[Arvanitaki et al.(2009)]{Arvanitaki} Arvanitaki, A. Dimopoulos, S., Dubovsky, S., 
Graham, P.; Harnik, R., Rajendran, S. 2009, Phys. Rev. D80.055011
\bibitem[Blumenthal \& Gould(1970)]{bg70} Blumenthal, G.R. Gould, R.J. 1970,
Rev. Mod. Phys. 42-2, 237
\bibitem[Chen et al.(2009)]{chen} Chen, C.R., Mandal, S.K., Takahashi, F. 2009,
(arXiv:0910.2639v2)
\bibitem[Diemand et al.(2007)]{diemand} Diemand, J., Kuhlen, M., and Madau, P. 2007, ApJ 657, 252
\bibitem[Eichler(1989)]{Eichler1989} Eichler, D. 1989, PRL 63, 2440
\bibitem[Eichler \& Maor(2005)]{ME05} Eichler, D., and Maor, I. 2005, (arXiv:astro-ph/0501096)
\bibitem[Essig et al.(2009)]{essig} Essig, R., Sehgal, N., Strigari, L.E. 2009, PRD 80, 223506
\bibitem[Franceschini et al.(1998)]{franceschini} Franceschini, A., Rodighiero, G., Vaccari, M.
2008, A\&A 487, 837
\bibitem[Grasso et al.(2009)]{lat-int} Grasso, D., Profumo, S., Strong, A.W., et al. 2009,
Astrop. Phys. 32, 140
\bibitem[Hooper et al.(2009)]{hooper09} Hooper, D., Stebbins, A.,
Zurek, K.N. 2009, PRD 79, 103513
\bibitem[Ishiwata et al.(2009)]{ishiwata} Ishiwata, K. Matsumoto, S., Moroi, T. 2009,
Phys. Lett. B 679-1, 1
\bibitem[Kappadath(1999)]{kapp} Kappadath, S.C. 1999, Ph.D. Thesis, University of New Hampshire
\bibitem[Kistler \& Siegal-Gaskins(2009)]{KS} Kistler, M.D., Siegal-Gaskins, J.M. 2009, arXiv:0909.0519
\bibitem[Kuhlen \& Malyshev(2009)]{km09} Kuhlen, M., \& Malyshev, D. 2009, PRD 79, 123517
\bibitem[Kuhlen et al.(2009)]{kuhlen07} Kuhlen, M., Diemand, J., Madau, P. 2007, ApJ 671, 1135
 \bibitem[Kuhlen, Madau, and Silk (2009)]{KMS} Kuhlen, M., Madau,P., \& Silk, J. 2009, Science, 325, 970).
\bibitem[Law et al.(2009)]{law09} Law, D.R., Majewski, S.R., Johnston, K.V. 2009, ApJ 703, L67
\bibitem[Moskalenko \& Strong(1998)]{ms98} Moskalenko, I.V., Strong, A.W. 1998, ApJ 493, 694
\bibitem[Navarro et al.(1996)]{navarro} Navarro, J.F., Frenk, C.S., White, S.D.M. 1996, ApJ 462, 563
\bibitem[Pohl et al.(2003)]{p03} Pohl, M., Perrot, C., Grenier, I., Digel, S. 2003, A\& A 409, 581
\bibitem[Pohl \& Esposito(1998)]{pe98} Pohl, M., Esposito, J.A. 1998, ApJ
507, 327
\bibitem[Porter \& Strong(2005)]{porter} Porter, T.A. \& Strong, A.W. 2005,
Proc. of the 29th ICRC, Pune, (arXiv:astro-ph/0507119)
\bibitem[Profumo \& Jeltema(2009)]{profumo09} Profumo, S., Jeltema, T.E. 2009,
JCAP 07, 020
\bibitem[Profumo(2008)]{profumo08} Profumo, S. 2008, arXiv:0812.4457v2
\bibitem[Sreekumar et al.(1998)]{sreekumar} Sreekumar, P., Bertsch, D.L.,
Dingus, B.L., et al. 1998, ApJ 494, 523
\bibitem[Strong et al.(2004)]{smr05} Strong, A.W., Moskalenko, I.V., Reimer,
O. 2004, ApJ 613, 956
\bibitem[Strong et al.(2000)]{strong00} Strong, A.W., Moskalenko, I.V., Reimer,
O. 2000, ApJ 537, 763
\bibitem[Tylka and Eichler (1987)]{TE87} Tylka, A.J and Eichler, D.
 (2007) U. Md. Technical Report,
\bibitem[Tylka (1989)]{Tylka1989} Tylka, A.J. 1989, PRL 63, 40
\bibitem[Zhang et al.(2009)]{zhang} Zhang, J. et al. 2009, PRD 80, 023007
\end{thebibliography}
\end{document}